\documentstyle[prb,aps,eqsecnum,floats]{revtex}

\begin{document}
\draft
\twocolumn
\input epsf

\title{Quantum mechanical analysis of the elastic propagation\\
of electrons in the Au/Si system: application to
Ballistic Electron Emission Microscopy 
}

\author{K. Reuter \cite{Erlangen}, P.L. de Andres}

\address{Instituto de Ciencia de Materiales (CSIC), \\
Universidad Autonoma de Madrid, E-28049 Madrid (Spain)}

\author{F.J. Garcia-Vidal, D. Sestovic, F. Flores}

\address{Departamento de Fisica Teorica de la Materia Condensada (UAM), \\
Universidad Autonoma de Madrid, E-28049 Madrid (Spain)} 

\author{K. Heinz}

\address{Lehrstuhl f\"ur Festk\"orperphysik,
Universit\"at Erlangen-N\"urnberg,\\ Staudtstr. 7, 91058 Erlangen (Germany)}

\maketitle

\begin{abstract}
We present a Green's function approach based on a LCAO
scheme to compute the elastic propagation of electrons 
injected from a STM tip into a metallic film. The obtained
2D current distributions in real and reciprocal space
furnish a good representation of the elastic component of 
Ballistic Electron Emission Microscopy (BEEM) currents.
Since this component accurately approximates the total
current in the near threshold region, this procedure
allows -- in contrast to prior analyses -- to take into
account effects of the metal band structure in the modeling
of these experiments. The Au band structure, and in particular
its gaps appearing in the [111] and [100] directions,
provides a good explanation for the previously irreconcilable
results of nanometric resolution and similarity of BEEM
spectra on both Au/Si(111) and Au/Si(100).
\end{abstract} 


\section{INTRODUCTION}

Ballistic Electron Emission Microscopy (BEEM)\cite{kaiser,mario}
is a new technique based on the Scanning Tunneling Microscope (STM). 
It has been primarily designed for the study of buried metal-semiconductor
interfaces, in particular for the investigation of the Schottky
barrier. The experimental setup consists of a STM injecting current
in a metallic film deposited on a semiconductor material. After propagation
through the metal, a fraction of these electrons still has sufficient
energy to surpass the Schottky barrier and may enter into the
semiconductor to be finally
detected as BEEM current. Using the tunneling
tip as a localized electron source gives BEEM its unparalleled power to
provide spatially resolved information on the buried interface, that can
additionally be related to the surface topography via the simultaneously
recorded tunneling current.

The energy of the electrons contributing to the final BEEM current depends
on the bias voltage between tip and metal, and is typically 1 to 10 eV above
the Fermi energy in the metal. For energies close to the threshold voltage
(the ones of primary interest for exclusive Schottky Barrier Height
(SBH) investigations), thin metallic layers, and low temperature, the
main contribution to the BEEM current stems from the elastic component,
which are electrons not having suffered losses from electron-electron 
and/or electron-phonon interaction. Only in this limit can the technique be
properly called {\em ballistic} \cite{cardiff}, and we shall concentrate in
this paper on the propagation of such electrons and their contribution to
the BEEM current.

The Au/Si interface has been one of the first systems investigated by BEEM
and has also proven to be a controversial one. Two important and apparently
irreconcilable results have been (i) the elegant demonstration of nanometric
resolution ($\approx 15${\AA}) after propagation through films of
as much as $100-150${\AA} \cite{milliken}, and (ii) the very similar results
obtained for the interfaces with Si(111) and Si(100) 
\cite{kaiser,schowalter,ludeke}
despite the strongly different k-space distribution of the projected Conduction
Band Minima (CBM) available for injection of electrons into the semiconductor.

The theoretical analysis of BEEM data is usually based on the so-called four
step model \cite{mario}: 1) tunneling from the tip to the surface,
2) propagation of the electrons through the metallic layer, 3) injection into
the semiconductor and 4) effects associated to various current-changing
processes in the semiconductor (e.g. impact ionization and/or creation of
secondary electrons). The last process becomes, however, only important at
rather high energies and can be safely neglected when concentrating on
the near threshold regime. The standard model used to extract SBH's from
the experiment is a simple quasi 1D semiclassical approximation using planar
tunneling theory and WKB approximation for step 1), free electron propagation 
plus
simple exponential attenuation for step 2) and the QM transmission coefficient
derived for a 1D step potential for step 3). This predicts a 5/2-power law for
the onset of the BEEM current above the threshold energy and has been used
extensively for the fitting of experimental data \cite{mario}.

Despite its obvious merits and especially its highly welcomed simplicity,
the standard model fails for more than a mere qualitative
explanation of experimental data. In particular, the above mentioned results
for the Au/Si system are not comprehensible within the framework of this
crude approximation. As has been pointed out recently \cite{prl}, the
major error leading to these discrepancies for Au/Si is the complete
neglect
of band structure effects inside the metal film. The free electron treatment
of the existing model predicts an electron beam propagating in normal direction
through the metal film, grown in [111] direction on either Si(111) or Si(100)
\cite{akgreen,oura}. On the other hand, it is a well known that Au shows a
band gap in just this direction \cite{ashcroft}, i.e. the Au band structure
clearly forbids electronic propagation normally through the film.
 
Unfortunately, the inclusion of band structure effects into a theoretical
model requires a much higher level of sophistication than for the previous
simple approximation based on E-space Monte-Carlo and
various parametrized processes. We therefore present in this paper a Linear Combination
of Atomic Orbitals (LCAO) scheme for the fully quantum mechanical computation
of the elastic component of the BEEM current: step 1) is included by coupling
the tip to a semi-infinite crystal via the Keldysh Green`s function formalism
\cite{keldysh}, whereas for step 2) we can take the Au band structure fully
into account via the corresponding Slater-Koster parameters
\cite{slater,constan}. This concept allows to compute the electron current
distributions in real and reciprocal space in any layer inside the metal film.

We will show in this paper, that already the detailed analysis of these
2D distributions immediately resolves the previously puzzling results
for Au/Si as a pure consequence of the band gap enforced sideward propagation
of the electron beam inside the metal film. This sideward beam is highly
focused thus explaining the obtained nanometric resolution, but 
it is also
dominated by electrons with rather high ${\bf k}_{\|}$-momentum parallel
to the interface, which allows similar injection into the projected Si CBM
for both Si(111) and Si(100) interfaces. Even though our proposed model is 
a little more demanding than the old simple one 
both from a theoretical and a computational
point of view, we believe that it allows to gain deeper insight into the
physics involved and that it represents a step towards a more careful
treatment of the BEEM process which ultimately will enable a much better
use of the obtained experimental data.

\section{THEORY}

Our model is based on a Hamiltonian written in a LCAO basis:

\begin{equation}
\hat H = \hat H_{T} + \hat H_{S} + \hat H_{I},
\label{hamilton}
\end{equation} 

\noindent
where $\hat H_{T} = \sum \epsilon_{\alpha} \hat n_{\alpha} + \sum \hat
T_{\alpha \beta} \hat c_{\alpha}^{\dagger} \hat c_{\beta}$ defines the
tip, $\hat H_{S} = \sum \epsilon_{i} \hat n_{i} + \sum \hat T_{i j}
\hat c_{i}^{\dagger} \hat c_{j}$ designates the metal sample and
$\hat H_{I} = \sum \hat T_{\alpha m} \hat c_{\alpha}^{\dagger} \hat c_{m}$
describes the coupling between the tip and the surface in terms of a
hopping matrix, $\hat T_{\alpha m}$ ($\hat n_{\alpha}$,
$\hat c_{\alpha}^{\dagger}$, and $\hat c_{\alpha}$  are number, creation
and destruction operators defined in the usual way). Note, that
greek indices indicate tip sites, whereas latin indices correspond to sites
inside the sample. The tight-binding parameters $\hat T_{i j}$ allow us to
fully take into account the sample band structure and can be obtained
within the empirical tight-binding framework by a fitting to {\em ab-initio} 
band
structures \cite{slater}. Tabulated values exist for all
elemental crystals, in particular for Au as used in the present
paper \cite{constan}. The hopping matrix between tip and sample
$\hat T_{\alpha m}$ is also expressed as a function of the different
atomic orbitals of atom $\alpha$ in the tip and atom $j$ in the sample.
Since BEEM experiments operate at high bias and rather low current,
where the tip-surface distance is already of the order
of $5-10$ {\AA}, only tunneling between s-orbitals needs to be considered
without loss of generality. The actual matrix element for this tunneling
between the s-orbitals is simply described by an exponential function,
as derived from the WKB approximation \cite{chen}. 
It is important that
$\hat T_{\alpha m}$ is localized in a small region of $m$ atoms close
to the tip. 

A Green's function formalism presents the important advantage of being free
of any adjustable parameters in the strictly elastic limit, where only an
arbitrarily small positive imaginary part $\eta$ is added to the energy, $E$,
necessary to ensure mathematical convergence:

\begin{equation}
\hat G^{R}(E) = \frac{1}{E - \hat H + i \eta}.
\label{green}
\end{equation}

\noindent
Although in this paper we are not interested in discussing inelastic effects,
which will form part of a forthcoming paper, it is worth mentioning that a
finite $\eta$ can be used to give some attenuation to the wavefield, 
mimicking inelastic effects that draw away current from the BEEM experiment. 
We shall see in the discussion of the k-space current distributions that 
a finite $\eta$ effectively
defines a coherence region of size approximately given by $\lambda_{c}
\approx { k_{F} \over \eta }$. Inside this region, quantum mechanical effects
are important, but outside of it quantum coherence is lost and intensities
rather than amplitudes should be added to compute the final wavefield.  

The system under investigation is out of equilibrium as soon as 
a bias between tip and sample is applied\cite{caroli}. In order to
retain the {\em ab-initio} advantage of a Green`s function formalism,
but still to be able to couple the tip to the sample, it is convenient
to use the Keldysh technique \cite{keldysh}, which essentially represents 
the generalization of many-body Green`s function theory to systems out of 
equilibrium \cite{landau}. Within this formalism, the current between two
sites $i$ and $j$ in the sample can be written as:

\begin{equation}
J_{ij} = \frac{e}{\pi \hbar} \int Tr \{ \hat T_{i j}
(\hat G_{ji}^{+-} - \hat G_{ij}^{+-}) \} dE.
\label{start}
\end{equation}

\noindent
The matrices $\hat G^{+-}$ are non-equilibrium 
Keldysh Green's functions
that can be calculated in terms of the 
retarded and advanced Green's functions
$\hat G^{R}$/$\hat G^{A}$ 
of the whole interacting tip-sample system.
We are interested only in the elastic
component of the current,

\begin{equation}
\hat G^{+-} = 
( \hat I + \hat G^{R} \hat \Sigma^{R} ) \hat g^{+-} 
( \hat I + \hat \Sigma^{A} \hat G^{A} ),
\label{eqgkeldysh}
\end{equation}

\noindent
where $\hat \Sigma^{R,A}$ is the coupling between tip and sample that can
be build up from the hopping matrices, $\hat T_{\alpha j}$, linking
atoms in both subsystems, while $\hat g^{+-}$ refers to the 
Keldysh Green's functions of the uncoupled parts before
any interaction has been switched on.

The retarded and advanced Green's functions of the interacting system 
can be further obtained from a Dyson-like equation that uses the Green's
functions of the uncoupled parts of the system,
$\hat g^{R}$/$\hat g^{A}$, and the
coupling between the surface and the tip, $\hat \Sigma^{R,A}$:

\begin{equation}
\hat G^{R,A} = \hat g^{R,A} + 
\hat g^{R,A} \hat \Sigma^{R,A} \hat G^{R,A}.
\label{eqgdyson}
\end{equation}

After some algebra, it is shown in Appendix A
that the current between two sites $i$ and $j$
in the metal can be obtained, in the lowest order of perturbation
theory with respect to the coupling between 
tip and sample, from the following formula:

\begin{eqnarray}
J_{ij}(V) = 
\frac{4e}{\hbar} \Im \int_{eV_o}^{eV} 
\! \!  \! \! Tr \! \! \! \sum_{m \alpha \beta n}  \! \! \! \left[
\hat T_{ij} \hat g_{jm}^{R} \hat T_{m \alpha} 
\hat \rho_{\alpha \beta}
\hat T_{\beta n} \hat g_{ni}^{A} \right] dE,
\label{jreal}
\end{eqnarray}
\noindent
where the integration is performed between the
Schottky barrier ($eV_o$) and the voltage ($eV$) applied 
between the tip and the sample. 
We notice that this equation shows
spectroscopic sensitivity because the Green's functions,
$\hat g^{R,A}_{jm}$, and the hopping matrices,
$\hat T_{m\alpha}$, depend on the energy, and also
through the upper limit in the integral.
The summation runs over tunneling active 
atoms in the tip ($\alpha$, $\beta$) and the sample 
($m$, $n$), 
$\hat \rho_{\alpha \beta}$ is the density of
states matrix: 
$( \hat g_{\alpha \beta}^{A} - \hat g_{\alpha \beta}^{R} ) = 
2 \pi i \hat \rho_{\alpha \beta}$.
The trace denotes summation over the atomic
orbitals forming the LCAO basis. 

Note, that the detour via the Keldysh technique has enabled us to 
arrive at an expression that only includes the (tabulated)
tight-binding parameters $\hat T_{ij}$ inside the sample, the hopping
elements $\hat T_{m \alpha}$/$\hat T_{\beta n}$ between tip and sample,
and the retarded and advanced equilibrium Green`s functions 
$\hat g^{R}$/$\hat g^{A}$ of the isolated tip and sample.

We shall describe the injection of electrons from the metal into the
semiconductor simply by matching the corresponding states. Therefore, we
shall need to compute the current distribution at the interface in 
reciprocal space. Working in a similar way as above for the real space,
we find the following expression for the current distribution in reciprocal
space defined now between planes $i$ and $j$:

\begin{eqnarray}
J_{ij}({\bf k}_{\|},V) \;&=&\; 
\frac{4e}{\hbar}  \;
\Im \int_{eV_o}^{eV} \; 
Tr \sum_{m \alpha \beta n} \left[ \;
\hat T_{ij} ({\bf k_{\parallel}}) \; 
 \hat g_{jm}^{R}({\bf k_{\parallel}}) \right. \nonumber \\ \;
& & \left.  \hat T_{m \alpha}({\bf k_{\parallel}}) \;
\hat \rho_{\alpha \beta}({\bf k_{\parallel}}) \;
\hat T_{\beta n}({\bf k_{\parallel}}) \;
\hat g_{ni}^{A}({\bf k_{\parallel}}) \right] dE
\label{reciprocal}
\end{eqnarray}
\noindent
where in this case the sum runs over layers in 
the tip ($\alpha$, $\beta$) and sample ($m$, $n$)  
that are involved in the tunneling process.
All quantities in eq. (\ref{reciprocal}) 
are the $k_{\parallel}$-Fourier 
transforms of the
corresponding objects in formula (\ref{jreal}). The total current 
between planes $i$ and $j$ can be obtained by summing
the current distribution $J_{ij}({\bf k}_{\|},V)$ as obtained in eq.
(\ref{reciprocal}) over the 2D Brillouin zone.

It is worth mentioning, that even though the total resulting current at
the given STM bias $V$ is obtained as an integral in energies down to
the SBH, the dominant contribution will be just the one at $V$ itself,
due to the strong exponential dependence of the tunneling coupling 
terms $\hat T_{\alpha j}$ with energy. This shall allow us to present
qualitative results displaying only this main contribution 
(calculated at the highest energy, $eV$), 
although we have additionally verified that essentially the same
effects occur at lower energies.

>From equation (\ref{jreal}) it is clear that the main objects of
interest that still need to be calculated are the retarded and advanced
equilibrium Green`s functions $g_{jm}^{R}$/$g_{ni}^{A}$ for a metallic surface.
We compute these quantities using a decimation technique \cite{guinea,lanoo},
which involves an iterative process allowing in each step to double in size
an existing slab (the so-called "superlayers" in 
Figure \ref{geometry}a). 
Assuming perfect periodicity in both directions parallel to the surface
one can easily calculate $\hat g_{11}^{(0)R}({\bf k}_{\|},E)$, the retarded 
Green's function of the surface superlayer which contains the tunnel-active
layers $m,n$. The layer doubling
of the slab is then repeated until its size is so large, that the two surfaces
are effectively decoupled. Hence, we obtain the Green`s function 
projected at the surface, $\hat g_{11}^{R}({\bf k}_{\|},E)$, and the 
{\it transfer matrix} of the system\cite{phscr}. 
With these two quantities it is 
possible to calculate the Green's function $\hat g_{j1}^{R}({\bf 
k}_{\|},E)$, 
representing the 
propagation of an electron of momentum ${\bf k}_{\|}$ and energy 
$E$ from the surface superlayer $1$  up to the layer $j$ inside the 
semi-infinite crystal.
The corresponding Green's function 
in real space is subsequently obtained
by Fourier transforming:

\begin{figure}
\epsfxsize=0.5\textwidth \epsfbox{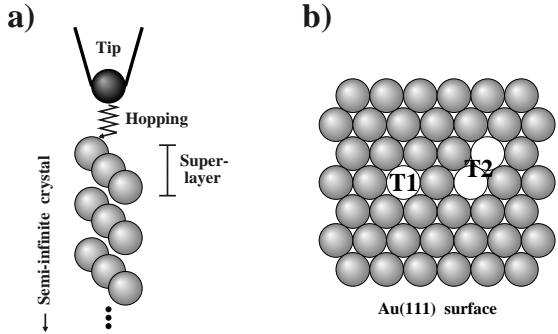}
\caption{Description of the geometry of the system. 
a) lateral view: tip coupled to the semi-infinite crystal, which is
composed of superlayers used for the decimation; b) Au(111) surface
depicting the two tip positions treated in the text: T1 directly on
top of one atom, T2 symmetrical bridge site between two atoms (atoms
in the surface to which tunneling is considered are whitened).
}\label{geometry}
\end{figure}

\begin{equation}
g_{j1}^{R} (r_{\parallel},E) =
\sum_{{\bf k_{\parallel}}} w_{k_{\parallel}}\;
g_{j1}^{R} ({\bf k_{\parallel}},E)
e^{ i {\bf k_{\parallel}} r_{\parallel} },
\end{equation}

\noindent
where the summation is performed over N special points \cite{rafa} in the 2D
Brillouin zone with respective weights $w_{k_{\parallel}}$. Finally, it is
straightforward, to obtain the corresponding advanced Green`s function
by transposing and conjugating:

\begin{equation}
\hat g_{1j}^{A} (E^{*}) = \left[ \; \hat g_{j1}^{R} (E) \; \right]^{*}
\label{advaret}
\end{equation}

Equations (\ref{jreal})-(\ref{advaret})
define our basic approach to the BEEM
problem.

Let us draw our attention to the full quantum-mechanical
calculations in real space based on equation (\ref{jreal})
and the decimation technique.
The total procedure outlined in this section enables the computation of 2D
current distributions at any layer inside the semi-infinite crystal.
Using eq. (\ref{jreal}), and calculating $J_{ij}(V)$ for all atoms $i$
inside a given layer,
 with current contributions from all their respective
neighbours $j$ in the layer above, we arrive at the current distribution
in real space. This simulates the BEEM current impinging on the
semiconductor after propagation through a metallic film with a 
thickness corresponding to the chosen layer $i$. Results of this type
will help us to understand which spatial resolution can be expected
in a BEEM experiment. Alternatively, the calculation of
$J_{ij}(\bf{k}_{\parallel},V)$
according to eq. (\ref{reciprocal})
provides the current distribution between layers $i$ and $j$
inside the 2D projected Brillouin zone. Since ultimately the electrons
need not only to have sufficient energy to enter the semiconductor as
BEEM current, but also to have the corresponding $\bf{k}_{\parallel}$
to get into the Si CBM, this type of plot will permit us to investigate
the onset and strength of the BEEM current for a certain type of interface.

When concentrating on the near threshold region, i.e. injection into
the bottom of the Si conduction bands, the latter may be approximated
by free electron-like paraboloids with appropriate
effective masses\cite{ashcroft}. In the 2D
projection on either the Au/Si(111) or Au/Si(100) interface this will
then lead to a number of ellipses inside the Brillouin zone, representing
areas through which transmission into the semiconductor is possible
\cite{guthrie}. To arrive at quantitative results, additionally a QM
transmission factor for each of the matched states has subsequently to be
applied. However, in the present paper we shall be concerned rather with
qualitative results apparent from the 2D distributions themselves, and
may thus leave the question of modeling the transmission factor for a
forthcoming publication \cite{future}. We feel that more physical insight
is gained by visually comparing how the current distribution in the metal
matches the available states in the semiconductor, than by presenting the
final BEEM current, which results from a summation over all
states with right matching conditions inside the 2D Brillouin zone. 
Therefore, in this work we
simply draw in the 2D Brillouin zone the Si CBM ellipses to aid the eye
identifying the relevant regions. 
We stress that the final summation tends to smooth details 
and, as the physically measurable
observable is only the $k_{\parallel}$-integrated $I(V)$ curve,
this might be one reason that helps to explain 
why the previous simple models achieve in
many cases decent fits to the data, even though it is quite obvious that
the underlying 2D current distribution is grossly inconsistent
with the metal band structure. 

\section{Real space results}

We apply our quantum-mechanical formalism to describe the elastic propagation
of electrons through a [111]-oriented Au metallic layer. Due to the large
lattice mismatch between Au ($a_{fcc} = 4.08${\AA}) and Si
($a_{diamond}=5.43${\AA})\cite{wyckoff}, the formed interface is of rather
poor quality. Although the growth of Au layers on Si has not yet been studied
extensively, there is evidence from Low Energy Electron Diffraction
and Auger studies, that suggest that Au films growing either on Si(111)\cite{akgreen} or
on Si(100)\cite{oura} form crystals oriented in the [111] direction after the
first four or five layers, which typically display heavy disorder, are
completed. From the nature of our results, that show the gradual build-up of
Bloch waves as a function of the distance to the surface, 
we deduce that the
likely disorder contained in four or five layers immediate to the interface,
does not provoke the automatic destruction of the formed pattern. We therefore
take the calculated 2D distributions in any given layer  of the ideal
semi-infinite crystal as a good approximation to the actual current arriving
at the Si semiconductor after propagation through a Au film of corresponding 
thickness and passage through the disordered interface region.

Before discussing our full quantum-mechanical results, 
it is worth analyzing
the semiclassical limit, which is based on Koster's approximation
to compute the bulk Green's functions\cite{koster} and an appropriate
symmetrization of waves reflected at the surface\cite{verde}. The central
point in this semiclassical approach mainly 
valid for thick layers is that
it admits a simple geometrical interpretation for the bulk propagators,
relating them to the shape of the constant energy surface (Fermi surface)
of the metal band structure: applying a stationary phase condition all
$\bf{k}$-values contributing to the final Green's function may be neglected
except one particular $\bf{k}_{0}$\cite{koster}:

$$
\hat g_{j1}^{R}(E) =
\sum_{\bf k} \; {e^{i {\bf k} ({\bf R_{j}} - {\bf R_{1}})}
\phi^{*}_{\bf k}(j) \phi_{\bf k}(1) \over E - E(\bf{k}) }
$$
\begin{equation}
\approx - { 1 \over \sqrt{
{ \partial^{2} \epsilon \over \partial k_{1}^{2} }
{ \partial^{2} \epsilon \over \partial k_{2}^{2} }
} }
e^{i
{
{\bf k_{0}} ( {\bf R_{j}} - {\bf R_{1}} )
\over
\mid {\bf R_{j}} - {\bf R_{1}} \mid 
}
}
\phi_{ {\bf k_{0}} }^{*} (j) \phi_{ {\bf k_{0}} } (1)
\label{oldg}
\end{equation}

\noindent
where $\partial^{2} \epsilon \over \partial k_{1}^{2}$,
$\partial^{2} \epsilon \over \partial k_{2}^{2}$ are the two principal
curvatures associated with the constant energy surface and
the eigenvectors $\phi_{\bf{k}_{0}}^{*}(j)$, $\phi_{\bf{k}_{0}}(1)$
are obtained diagonalizing the bulk tight-binding Hamiltonian $H$.

\begin{figure}
\epsfxsize=0.5\textwidth \epsfbox{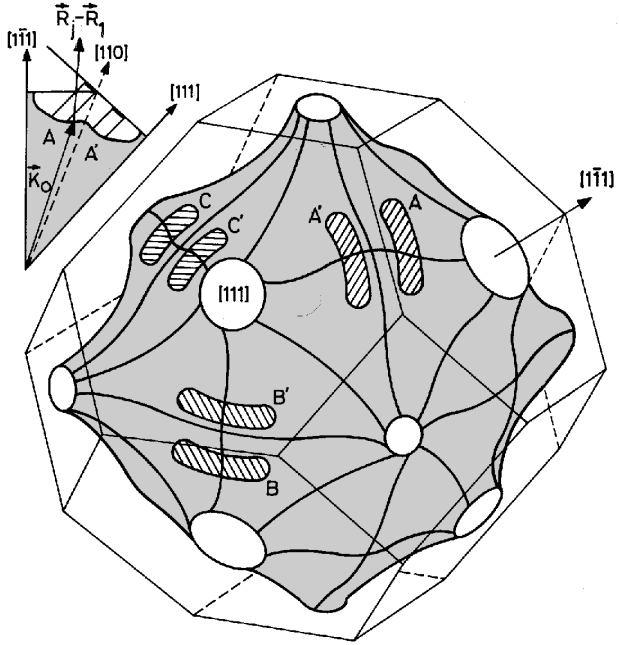}
\caption{Shape of the Au constant energy surface at 1eV above $E_F$.
The shaded ellipses mark the regions contributing most to
the projected 2D current distributions. The inset shows a
schematic description for the relation between a real space
propagation direction $(\bf{R}_{j} - \bf{R}_{1})$, and the
relevant wave vector $\bf{k}_{0}$ defined in the semiclassical
approximation.
}
\label{fermisurf}
\end{figure}

For a given propagation direction $(\bf{R}_{j} - \bf{R}_{1})$,
$\bf{k}_{0}$ is the vector linking the $\bf{k}$-space origin to the point
on the constant energy surface (defined by the energy $E$) whose
perpendicular to the surface is parallel to the initially chosen
direction in real space (see inset in 
Fig. \ref{fermisurf}). 
This vector perpendicular
to the constant energy surface at $\bf{k}_{0}$ represents the
group velocity 
of electrons being propagated in the metal, 
$\bf{v}_{g}(\bf{k}_{0}) = \nabla_{\bf{k}} E(\bf{k})$,
that 
needs not to be parallel to $\bf{k}_{0}$ when the metal band
structure deviates from the free electron limit. Eq. (\ref{oldg})
reveals that the current intensity is strongly enhanced in directions
corresponding to rather flat parts of the constant energy surface,
where the principal curvatures
$\partial^{2} \epsilon \over \partial k_{1}^{2}$ and
$\partial^{2} \epsilon \over \partial k_{2}^{2}$ in the denominator
tend towards zero (note, that at the extremal points,
where the second derivatives are exactly zero, the next term in Koster's
expansion is needed). Hence, already a simple visual analysis of the
constant energy surface's shape may provide an intuitive understanding where 
major
ballistic current contributions are to be expected when including band
structure effects.

As pointed out previously\cite{prl}, for the analysis of
electron propagation in Au layers at typical BEEM energies, it is crucial
to notice that for voltages larger than $\approx 1.0$V necks develop 
in the [100] directions, which are similar to the well known necks in
the [111] directions,
cf. 
Fig. \ref{fermisurf}. Identifying the flat parts on this surface and
their corresponding real space propagation directions, we were able to make
a qualitative prediction of the current distributions in real and reciprocal
space after propagation through a thick Au film\cite{prl}. 

In Fig. \ref{fermisurf}, 
we have drawn the Au constant energy surface for 1 eV
above $E_{F}$, and the regions (dashed areas) giving the
major contribution to the semiclassical BEEM current
in k-space. One of these areas is also shown in the
figure inset: the reason why the k-space current
distribution maximum is associated with the A-region,
and not with the apparently equivalent A'-region, is
their different orientation w.r.t. the (111)-direction
(see. Figs. 6 and 8). It should be kept in
mind that the dashed areas of 
Fig. \ref{fermisurf} define the
momentum and velocity components of the electrons 
contributing most to the BEEM-current: the group velocity
defining the semiclassical propagation in real space.
Since most BEEM modeling up to now has been performed using E-space 
Monte-Carlo techniques, 
it may be interesting to notice, that apart
from the neck regions where no propagation is possible, 
the Au constant energy
surface remains spherical to a good approximation 
(Fig. \ref{fermisurf}):
the group velocity $\bf{v}_{g}(\bf{k}_{0})$
perpendicular to the surface and the connecting k-vector $\bf{k}_{0}$
do not diverge by more than $\approx 20-30^{\circ}$ for most directions
of interest outside the necks. When properly suppressing the forbidden 
propagation directions corresponding to the necks, a simple application
of k-space Monte-Carlo simulation should thus be possible, resulting in
analogous results to ours in the purely ballistic limit.
This allows the simulation of inelastic interactions
in an approximate but simple way through the inclusion
of the appropriate k-space scattering cross sections, 
reaching the interesting conclusion that those effects
do not significatively degrade the elastic predicted
resolution in real space\cite{msmf}.

The presented semiclassical approach helps to better understand
the numerical results in terms of the geometrical interpretation. 
However, we have also at hand
the quantum mechanical decimation technique for the computation of the
Green`s functions, which allows us to precisely calculate these distributions
and also not only for thick films, but for any thickness. 
Since we are only interested in the qualitative 
evolution of injected current
in a [111]-oriented Au film, we start with a simplified tunneling geometry,
in which hopping is only permitted between one tip atom $0$ and one sample
atom $1$ positioning the tip directly on top of the latter 
at $5.0$ {\AA } distance, cf. 
Fig. \ref{geometry}b. More realistic tunnel geometries
allowing hopping to more atoms lead to essentially the same results: one
advantage of properly taking the metal band structure into account is that
the results are not that much dependent on the initial tunneling conditions 
as long as metallic films of a certain thickness are involved. 
This is a
strong difference between our theory and the standard ballistic E-space
Monte-Carlo approach. The k-space distribution arriving at the semiconductor
under the assumption of free electron propagation
is inappropriate for both very thick and
very thin layers: (i) it is physically unreasonable to assume that what 
impinges on the interface (after a long propagation through the metal) is
just the initial tunneling distribution, and (ii) on the other hand,
when the film is thin enough for the details of tunneling to become
important, the crude planar theory based on WKB should not be applied.
On the contrary, in all the examples presented in this paper (where we have
deliberately avoided ultra thin layers), tunneling simply provides a starting
configuration, whereas the final current distribution at the interface is
dominated by the preferred propagation directions dictated by the metal band
structure. In any case, should the initial tunneling distribution become
important for some particular application, our formalism would permit
a more realistic point of view for that part of the problem.

\begin{figure}
\epsfxsize=0.5\textwidth \epsfbox{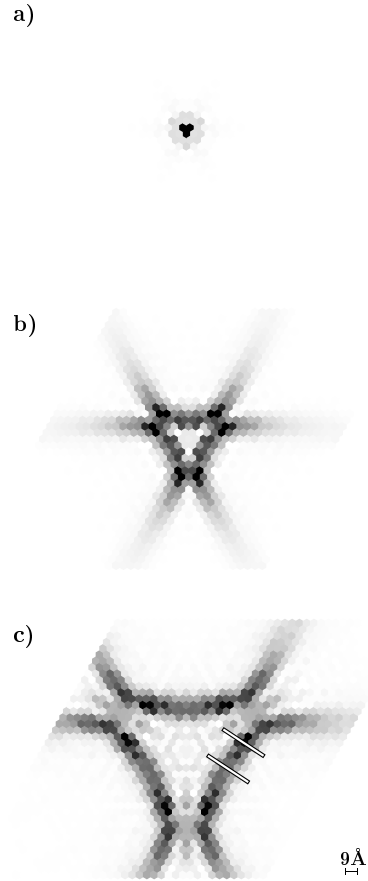}
\caption{Real space BEEM current distributions for Au(111) after
injection of current in one atom in the first layer (located in the
center of the drawn layer). The tip is positioned at $5.0${\AA} height
on top of the active atom. Parameters: $V=E_F+1$eV, $\eta=0.1$eV.
Distribution in 
a) 2nd layer (2.35{\AA}), b) 10th layer (21.19{\AA}), and
c) 25th layer (56.51{\AA}). 
Each dot represents one atom in the corresponding
layer and the grayscale indicates 
the amount of current passing through the atom:
black for maximum current to white for zero current.
}\label{realspace1}
\end{figure}

This discussion is illustrated in 
Fig. \ref{realspace1}, where the
resulting real space current distribution after electron propagation through 
more and more layers is depicted. As electron energy we chose 1 eV above the
Fermi level of the metal.
2971 special k-points were used inside the 2D Au Brillouin
zone to suppress possible aliasing effects in the involved Fourier transforms
to real space. In the second layer, very near to the surface (Fig.
\ref{realspace1}a), the current is still strongly concentrated in the atoms
closest to the location in the surface layer in which the current was
injected. This behaviour, where the current still propagates in all
directions and where the resulting distribution is basically a 
consequence of the crystal geometry and nearest neighbour hopping, can be
found down to about the fifth layer. Then, however, an interesting
change occurs, 
which is related to the gradual formation of a Bloch wave inside the crystal:
propagation becomes only possible in directions in accordance with the metal
band structure. The immediate consequence of the already mentioned band gap
of Au in the [111] direction is that the evolving beam has to make way
sidewards. This becomes obvious in Fig. \ref{realspace1}b, where the
opening up of the injected beam can already be observed with very little
current remaining in the central plane atoms lying in the [111] forward
direction.

A further effect of the band structure is the formation of narrowly focused, 
Kossel-like lines corresponding to preferred propagation directions. 
After propagation through a larger number of layers, a steady state is
reached and distributions for consecutive layers differ only by the natural
spreading of the resulting triangle as the electrons follow these preferred
propagation directions of the band structure, cf. 
Fig. \ref{realspace1}c. 
These directions have a polar angle of about $30^{o}$ with respect to the
normal direction, and are in good agreement with our previous semiclassical
analysis\cite{prl} (see Fig. \ref{fermisurf}). 
Note, how different the resulting pattern in each layer
is from the prediction of the simple, free electron model: there the
propagation would be solely dictated by the $\bf{k}_{\parallel}$-spreading
of the initial tunneling distribution resulting in a current cone centered
around the [111] direction. Consequently, the real space distribution 
in any layer parallel to the surface would be a filled circle 
(or rather a filled triangle, considering the symmetry of a fcc (111) crystal
\cite{cowley}) with the major current contributions directly in the middle
corresponding to normal propagation through the metal film.

\begin{figure}
\epsfxsize=0.5\textwidth \epsfbox{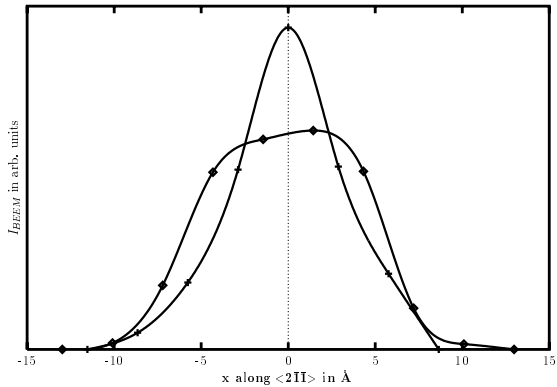}
\caption{Intensity profile along two perpendicular cuts through one of the
Kossel-like lines (the locations of the cuts are marked in Fig.
\ref{realspace1}c by white lines).}
\label{rprofile}
\end{figure}

The deeper the chosen layer, i.e. the larger the thickness of the
experimental film, the more spread over a larger triangle the current
would be. The simple application of the uncertainty principle to the 
tunneling process  
plus a free electron model for the metal \cite{milliken,schowalter}
 predicted a 
BEEM resolution for relatively thick $100-150${\AA} Au films of at best
$\approx 100${\AA}. On the contrary, the experiment by
Milliken et al. \cite{milliken} finds typical resolutions for such films 
on both Si(111) or Si(100) of about $15${\AA}. In that experiment, sharp
SiO${}_2$ steps are created on both Si substrate types and subsequently
covered by $\approx 150${\AA} Au films. BEEM images of the step riser, 
when the tip crosses from above a part corresponding to the impenetrable
SiO${}_2$ to a part, where BEEM injection into the Si is possible, give
a direct reflection of the size of the electron beam at the interface.
The typical $15$ {\AA } found are absolutely incompatible with the
standard ballistic, free particle propagation used in the simple model
and are difficult to justify even if parametrized
electron-electron/electron-phonon interaction is taken into
account\cite{milliken,schowalter}. On the other hand, we believe that the
observed formation of narrowly focused Kossel-like lines as caused by the Au band
structure may already explain the experimentally obtained nanometric
resolution in the purely elastic limit. As can be seen in Fig.
\ref{realspace1}c, the width of these lines carrying the BEEM current is
typically 3-4 atomic distances. This is better appreciated in Fig.
\ref{rprofile}, where cuts perpendicular through these lines at two different
positions were performed and the intensity vs. location is depicted. 
The derived width of $\approx 10${\AA} is in very good agreement with the
experimentally observed value and is clearly distinct from the value derived 
from free electron theory, which deviates by a complete order of magnitude.

A naive interpretation of this result seems to imply that multiple
images of interface objects should be observed superimposed, 
since several such Kossel-lines are present in our distribution. A tip scan
would sweep each of these lines across the interface object, each time
leading to a focused signal. It should, however, be born in mind, that
so far we have used unrealistically symmetric conditions for the current
injection process: the on top tip site used, allows the current pattern
to display the full threefold symmetry of the fcc crystal\cite{cowley} 
and hence predicts the existence of three equivalent Kossel-like lines. During
the scan, less symmetric tip sites are on the contrary involved, reducing
logically the resulting overall symmetry of the current distribution. To
verify this, we repeated the calculation, this time using a bridge site
for the tip and allowing tunneling to both closest atoms on the surface, cf.
Fig. ~\ref{geometry}b.

\begin{figure}
\epsfxsize=0.5\textwidth \epsfbox{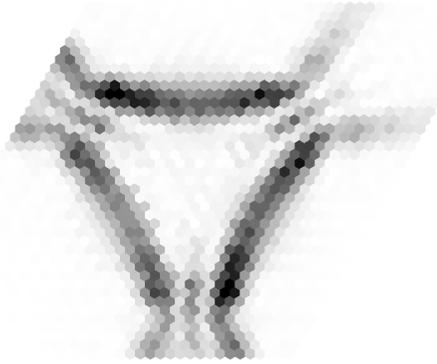}
\caption{Real space BEEM current distribution as in Fig. \ref{realspace1}c,
but injecting current under lower symmetry conditions: the tip is located in
a bridge site at a height of 5.0{\AA}.}
\label{realspace2}
\end{figure}

Several conclusions can be drawn from the result shown in Fig.
\ref{realspace2}, which is for the limiting case after propagation through
already a thick film. First, as expected, the overall symmetry of the pattern
has reduced from threefold to twofold, with one of the three Kossel-like
lines decreasing in intensity. Second and most importantly, the shape of the
distribution has essentially not changed. This is the confirmation of the
introductory statement made at the beginning of this section: 
the inclusion of band structure effects reduces considerably the influence of
the initial tunneling distribution. The preferred directions of propagation
are exclusively dictated by the band structure, whereas it is only the
population of these directions with electrons that depends on the actual
tunneling process. In this respect, we again observe in Fig. \ref{realspace2} 
the $\approx 30^{o}$ off-normal, sideward propagation along focused lines, 
of which now only the two corresponding to the chosen tunnel symmetry are
mainly populated.

As we are interested in highlighting the relevant physical phenomena, we
have introduced convenient approximations: one atom tip, spherical s-wave
tunneling and a perfect substrate without any tilt, nor defects on the
surface. In turn, the model neglects a number of factors that necessarily
would result in a lower symmetry than the one existing in a realistic
experimental setup. However, it is plausible
to accept that different factors
concur in the experiment to degrade symmetry, and will eventually lead to
the selection of only one of the possible Kossel-like lines, whose dominant
contribution will then be responsible for the observed BEEM image with its
nanometric resolution. On the other hand, none of the experimental
influences can change the metal band structure itself (as long as there are
at least crystal grains of the order of 100{\AA}), so that the formation of
focused Kossel-like lines will be very similar to the one seen in our idealized
model calculations.

\section{Reciprocal space results}

\begin{figure}
\epsfxsize=0.5\textwidth \epsfbox{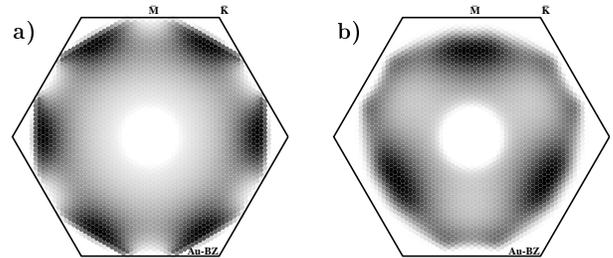}
\caption{Reciprocal space BEEM current distributions in the 2D interface
Brillouin zone. a) sixfold symmetry inside the coherence region,
$\eta=0.001$ eV b) threefold symmetry far outside the coherence region,
$\eta=0.1$ eV (2971 special k-points used in the 2D Brillouin zone,
$E=E_F+1$ eV; black dots represent high current, white zero current.
A quadratic grayscale has been applied to emphasize the current changes due
the symmetry crossover).}
\label{kspace1}
\end{figure}

In order to understand how the current distribution in the metal matches the
available states in the semiconductor, we now proceed to calculate the 2D
distributions in reciprocal space using expression (\ref{reciprocal}).
Fig. \ref{kspace1}a shows the result obtained inside the Au Brillouin zone, 
again for the symmetrical on top tip site, cf. Fig. \ref{geometry}b, 
and for an electron energy of $E=E_F+1$eV, which is still near
the threshold region for Au/Si. As a direct reflection of the Au [111] band gap, 
which projects onto $\bar{\Gamma}$ in the center of the hexagon, the
distribution displays a ring-like structure, since there are no states to
carry the current in the $\bar{\Gamma}$ vicinity. It is again interesting
to notice, how different this pattern is from the free electron prediction,
that would inject the main current with $\bf{k}_{\|}=0$ just into
$\bar{\Gamma}$. In reality, complete reflection occurs for tunnel electrons
with $\bf{k}_{\|}=0$, permitting only injection into the Au(111)
states of rather high $\bf{k}_{\|}$-momentum.

We would like to stress that the use of the decimation technique, 
which was derived from renormalization group techniques \cite{guinea},
permits the fully quantum mechanical calculation of both required Green`s
propagators $\hat g_{j1}^{R}$/$\hat g_{1i}^{A}$ of the semi-infinite slab.
The resulting BEEM current is consequently also fully quantum mechanical.
It is therefore not surprising that the distribution shown in Fig.
\ref{kspace1}a possesses a sixfold symmetry, which relates to the 
symmetry of the [111] projected density of states to which $+\bf{k}$ and
$-\bf{k}$ states contribute equally \cite{stiles}.

This is at variance with a semiclassical distribution, where only
$\bf{k}$-vectors representing propagation towards the interface must be
considered. These waves are classically separated from those that propagate
in the opposite direction and the total vector field has thus the same
symmetry as the constant energy surface, which in turn reflects the
symmetry of the lattice, i.e. threefold for [111] directions in a fcc material
\cite{ashcroft}. Hence, our previous semiclassical approach yielded
threefold symmetric BEEM current distributions \cite{prl}.

As has already been mentioned in section II, the use of a finite
self-energy $\eta$ in the construction of the Green`s functions (\ref{green})
puts some limits on the coherence length of our quantum mechanical
computation. In the purely elastic limit, corresponding to a 
theoretical $\eta\rightarrow0$ in eq. (\ref{green}), the obtained distribution 
would not change at all after propagation through an infinite number of
layers. In the ideal crystal, there are simply no processes enforcing
$\bf{k}_{\parallel}$-changes of the elastic electrons after their injection.
Numerically, we are however forced to choose a small, but finite $\eta$ to
ensure convergence. The result presented in Fig. \ref{kspace1}a was hence
obtained for $\eta = 0.001$ eV, which is very small compared to the energy
$E=1$ eV considered, and corresponds to a coherence length of several thousand
{\AA}ngstr{\o}m. No significant changes of the distribution can be observed
during the propagation through many layers. Since in this case we are always
well inside the coherence region, the symmetry is sixfold, as it should be
for a quantum mechanical calculation.

\begin{figure}
\epsfxsize=0.5\textwidth \epsfbox{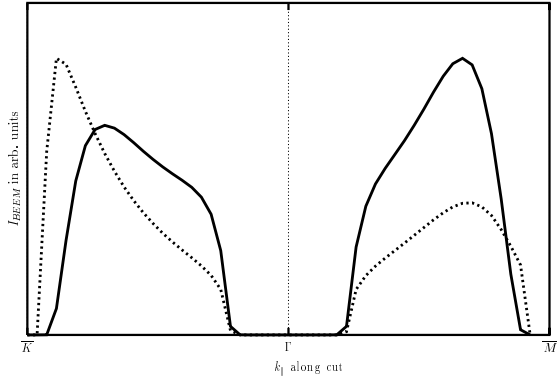}
\caption{Intensity profile along the high symmetry line
$\bar{K}-\bar{\Gamma}-\bar{M}$ of Fig. \ref{kspace1}. The solid line
represents current far outside the coherence region (cf. Fig. \ref{kspace1}b),
the dotted line current inside of it (cf. Fig. \ref{kspace1}a). The different
$\eta$'s lead to a different amount of damping in the currents and hence
different absolute scales. To enable better visual comparison, the currents
have therefore been renormalized to the same maximum value.}
\label{kprofile}
\end{figure}

On the other hand, we can considerably reduce the coherence length by using a 
much larger value for the optical potential, say $\eta=0.1$eV. There would
then exist a regime outside the coherence region where interference 
terms do not play a significant role any more. 
This should allow us to recover 
the semiclassical results. Fig. \ref{kspace1}b shows the obtained
distribution for such a large $\eta$ and for the 30th layer in the
crystal, i.e. well outside the coherence region. It is most gratifying that
not only the expected threefold symmetry appears, but that also the actual
pattern matches perfectly with our semiclassical prediction, 
that was then only derived
schematically, cf. Fig. 3 of ref. (\onlinecite{prl}). 
Moreover, it is interesting
to notice how our Green`s function calculation permits to reproduce the
involved change in symmetry: even for a large $\eta$, propagation in the
first layers is still inside the small coherence region and thus quantum
mechanical effects are observed to produce the corresponding sixfold symmetry. 
During the propagation through more and more layers, the quantum coherence is 
gradually lost as is the symmetry, which progressively approaches its
threefold limit. This allows to study the crossover between the quantum and
the semiclassical domain and provides another example of how a quantum
system, under the influence of friction, becomes classical by a decoherence
process \cite{zurek}.

Note, that the change in symmetry implies also a change of the detailed current
distribution itself: the maxima of the semiclassical pattern appear on the
lines $\bar{\Gamma}-\bar{M}$, whereas the quantum mechanical maxima lie along
the directions $\bar{\Gamma}-\bar{K}$. This is more obvious in Fig.
\ref{kprofile}, where an intensity cut along the symmetry lines
$\bar{K}-\bar{\Gamma}-\bar{M}$ is presented: the relative weight 
between currents along $\bar{K}-\bar{\Gamma}$ and $\bar{\Gamma}-\bar{M}$
respectively is inversed between the two types of calculations,
which is a reflection of the symmetry change. It is interesting to notice that
the evolution from the quantum symmetry to the classical one is gradually
build up, and no sudden change between both regimes is found. Our
calculations showed that the observed difference in reciprocal space does
not significantly affect the beams in real space (where the symmetry must
always be threefold), but it could nevertheless in principle affect the
I(V) current injected through the projected ellipses into the semiconductor.
However, because of the gradual crossover, no dramatic effects are to be
expected, unless one could experimentally break the time-reversal
symmetry suddenly (e.g. by application of a magnetic field).

\begin{figure}
\epsfxsize=0.5\textwidth \epsfbox{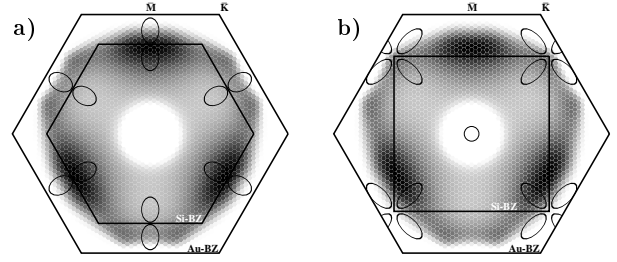}
\caption{Matching of the obtained semiclassical k-space BEEM current
distribution with the available CBM states in the semiconductor, 
which are approximated by projected parabolic bands. 
Current can only enter into the Si through the area enclosed by the ellipses
($E=E_F+1$eV). a) Au/Si(111), $E_{SBH} = 0.85$eV b) Au/Si(100),
$E_{SBH} = 0.82$eV. Note, that the different lattice parameters of Au and
Si require remapping of the Si ellipses inside the larger Au Brillouin zone.
$\bar{\Gamma}-\bar{M}$ corresponds to the [101] direction in k-space}
\label{remapping}
\end{figure}

As a final point we address the actual transmission of the current into the
semiconductor states. The chosen energy $E=E_F+1$eV is still in the near
threshold region ($E_{SBH}(Au/Si(111)) \approx 0.85$eV,
$E_{SBH}(Au/Si(100)) \approx 0.82$eV \cite{mario}) justifying to approximate
the Si CBM by nearly free electron ellipsoids. Since Si is an indirect
semiconductor, these ellipsoids are not located around $\Gamma$ in the
Brillouin zone, but ca. 85\% in direction $\Gamma-X$ \cite{ashcroft} and
hence project differently onto the [111] and [100] directions
\cite{mario,guthrie}. The resulting ellipses for both orientations are drawn
in  Fig. \ref{remapping}a and \ref{remapping}b. Note, that since we are
calculating inside the larger Au Brillouin zone, those ellipses belonging 
to higher Si Brillouin zones, but still within the first Au one, have to be
considered for current injection as well. The principal difference for 
both Au/Si(111) and Au/Si(100) interfaces is that in the latter case Si
ellipsoids project directly onto $\bar{\Gamma}$, which is where the simple
free electron picture puts the maximum current. Since very little current
carried by electrons with high $\bf{k}_{\parallel}$-momentum is predicted in
this type of model, a considerable difference in the onset and absolute
magnitude of the BEEM current was originally anticipated between both
interfaces \cite{mario,schowalter}. On the contrary, the actual,
experimentally observed spectra were highly similar
\cite{kaiser,schowalter,ludeke}. Since this result was irreconcilable
with the ballistic free electron theory, a variety of processes ranging from
more isotropic tunnel distributions \cite{milliken} over strong elastic
electron-electron interaction \cite{schowalter} to ${\bf k}_{\|}$-violation
at the non-epitactic interfaces \cite{ludeke2} were proposed in an extensive
debate in the literature. All these processes aimed at providing
$\bf{k}_{\parallel}$-momentum to the current distribution to allow
injection into the off-normal [111] ellipses, but ran into considerable
trouble through the simultaneous, inherent loss of resolution \cite{milliken}.

On the other hand, the ring-like current pattern predicted by our band
structure calculation is dominated by high $\bf{k}_{\parallel}$-momenta,
(cf. Fig. 8) 
but without
loss of resolution, as we have seen in the last section. The total reflection of the
injected electrons around $\bar{\Gamma}$ due to the projected Au band gap
renders the existence or non-existence of a Si ellipse in that region
completely irrelevant, since no current can enter through it into the
semiconductor in any case, and the current is therefore forced to enter via
any of the off-normal ellipses for both Au/Si(111) (Fig. \ref{remapping}a)
and Au/Si(100) (Fig. \ref{remapping}b). The ring-like distribution may
in zeroth order be well described as azimuthally symmetric, making it
highly plausible why the BEEM current shows similar onset and magnitude:
the exact location of the off-normal ellipses plays only a minor role. 

\begin{figure}
\epsfxsize=0.5\textwidth \epsfbox{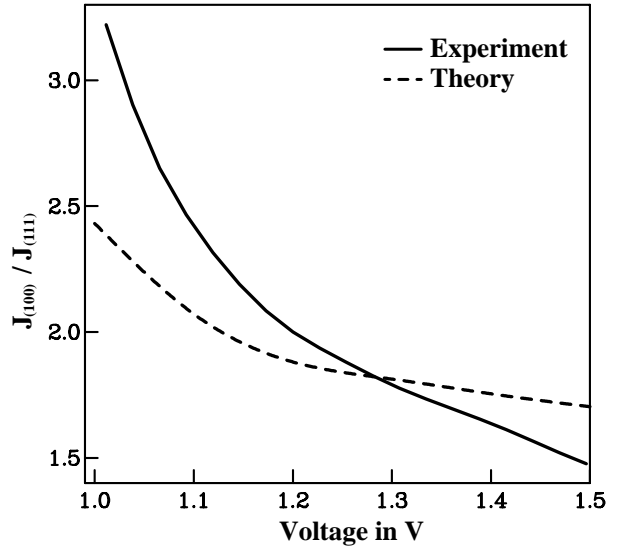}
\caption{
Ratio of current injected in Si(100) and Si(111) after
propagation through thin Au films of approximately equal thickness
(100 \AA\ Au/Si(100) and 75 \AA\ Au/Si(111), chosen to
compare with available experimental data). Elastic electron propagation
and T=1 are used in the theory (dashed line); experimental data taken
from refs. (\cite{bell,bell2}) (solid line).
}
\label{111vs100}
\end{figure}

Finally, we quantify this result by comparing the actual current injected
into the (100) and (111) silicon orientations. As mentioned in the
beginning, we shall not get at this stage into the separate problem of
computing a suitable transmission coefficient to keep our discussion as
simple as possible. Hence, we simply use $T=1$, which is justified because
we shall only present the ratio between the current injected into both
silicon faces $J_{100}/J_{111}$, where factors not very sensitive to a
particular silicon orientation cancel out. In particular, this is the case
for the transmission coefficient $T(E,\bf{k}_{\|})$ as can easily be
checked with the crude 1D step-barrier approximation\cite{mario}. Similarly,
we do not need to consider inelastic effects, that are anticipated to affect
equally electrons approaching the Si(100) or the Si(111) surface, and
should in general be small when considering thin metal films. The
ratio shown in Fig. \ref{111vs100} compares well with the experimental
ratio obtained from refs. (\onlinecite{bell,bell2}), reproducing
not only the correct order of magnitude, but also the overall trend with
energy. The largest difference of $\approx$30\% is seen in the low voltage
region where experimental currents are weak and difficult to measure, and
where the total current depends most sensitively on the exact value of
the SBH for both orientations, whereas we have made no effort in fitting
these values, but have simply taken averages over the big number of
existing experimental values (which differ easily by 0.1 eV). It is hence
not surprising that the agreement in the lowest 0.2 eV above the SBH
is not as convincing as it is at the higher voltages. Taking finally
into account the crudeness of the applied model where we have deliberately
avoided the fitting of any parameter to focus on physical insight,
we believe that the presented ring-like distribution provides a
satisfactory and intuitive explanation of the observed effect.

\section{Conclusions}

In conclusion, we have presented a theoretical model that allows the
calculation of the elastic contribution to the BEEM current in the near
threshold range. Based on a LCAO scheme and a Keldysh Green`s function
formalism, 2D distributions in real and reciprocal space may be computed
in any layer of a semi-infinite crystal after current is injected from a
tip atom. Subsequent matching with the semiconductor states permits to
extract the actual BEEM current, although the main emphasis in the present
paper has been on the qualitative understanding that can already be
obtained from the 2D patterns themselves. The model allows for the first time 
to fully take into account the influence of the metal band structure in the
BEEM  process, and possesses the further advantage of being free of any
adjustable parameter in the strictly elastic limit.

The application to the system Au/Si shows a variety of consequences of the Au
band gap in the [111] direction. In real space, the formation of narrowly
focused Kossel-like lines and a sideward beam propagation is observed, which
may explain the experimentally obtained nanometric resolution. In reciprocal
space, a symmetry change between quantum mechanical and semi-classical regime
can be related to the gradual breaking of quantum coherence. The actual
pattern in k-space has a ring-like shape, which calls for current injection
via the off-normal Si ellipses for both Au/Si(111) and Au/Si(100). Hence,
the sole inclusion of band structure effects achieves to explain the nanometric
resolution and the similarity of BEEM spectra on both Si orientations on
the same footing in the purely elastic limit -- without any adjustable
parameter and without the necessity to invoke any further scattering process.

\section*{Acknowledgments}

K.R. is grateful for financial support from SFB292 (Germany).
We also acknowledge financial support from the Spanish CICYT under
contracts number PB92-0168C and PB94-53.

\section*{APPENDIX A}

We start from formula (\ref{start}) and the Keldysh Green's functions
defined in (\ref{eqgkeldysh}). The coupling term 
$\hat \Sigma^{R,A}$
can be expressed as a function of the real hopping matrices,
$\hat T_{\alpha m}$, that link tunneling active atoms in the tip ($\alpha$) with 
the
corresponding ones in the sample ($m$).

Using equation (\ref{eqgkeldysh}), 
$\hat G_{ji}^{+-}$ and  $\hat G_{ij}^{+-}$ can
be written as:

\begin{eqnarray}
\hat G_{ji}^{+-} & = & g_{ji}^{+-} +
\sum_{\alpha m} \left[\hat G^{R}_{j\alpha} \hat T_{\alpha m}
g_{mi}^{+-}+
\hat g_{jm}^{+-} \hat T_{m \alpha} \hat G^{A}_{\alpha i}\right]+
\nonumber \\
& & \sum_{m \alpha \beta n} \hat G^{R}_{jm} \hat T_{m \alpha}
\hat g_{\alpha \beta}^{+-}
\hat T_{\beta n}
\hat G_{ni}^{A}
\end{eqnarray}

\begin{eqnarray}
\hat G_{ij}^{+-} & = & g_{ij}^{+-} +
\sum_{\alpha m} \left[\hat G^{R}_{i\alpha} \hat T_{\alpha m}
g_{mj}^{+-}+
\hat g_{im}^{+-} \hat T_{m \alpha} \hat G^{A}_{\alpha j}\right]+
\nonumber \\
& & \sum_{m \alpha \beta n} \hat G^{R}_{im} \hat T_{m \alpha}
\hat g_{\alpha \beta}^{+-}
\hat T_{\beta n}
\hat G_{nj}^{A}
\end{eqnarray}

The Keldysh Green`s function $\hat g^{+-}$
of the uncoupled
(and hence equilibrium) system can be expressed as a function of
the advanced and retarded Green's function of the uncoupled parts $\hat
g^{R}$/$\hat g^{A}$ and
the Fermi distribution
functions of the tip, $f_{T}(E)$, and sample, $f_{S}(E)$:

\begin{eqnarray}
\hat g_{mi}^{+-} \;&=&\; f_S \left( \hat g_{mi}^{A} -
\hat g_{mi}^{R} \right)\nonumber \\
\hat g_{jm}^{+-} \;&=&\; f_S \left( \hat g_{jm}^{A} -
\hat g_{jm}^{R} \right)\nonumber \\
\hat g_{\alpha \beta}^{+-} \;&=&\; f_T
\left( \hat g_{\alpha \beta}^{A} -
\hat g_{\alpha \beta}^{R} \right)\\
\hat g_{mj}^{+-} \;&=&\; f_S \left( \hat g_{mj}^{A} -
\hat g_{mj}^{R} \right)\nonumber \\
\hat g_{im}^{+-} \;&=&\; f_S \left( \hat g_{im}^{A} -
\hat g_{im}^{R} \right), \nonumber
\end{eqnarray}

Using these last equations we obtain:

\begin{equation}
\hat G_{ji}^{+-} - \hat G_{ij}^{+-} =
\hat g_{ji}^{+-} - \hat g_{ij}^{+-} +
f_{T} U_{T} + f_{S} U_{S}
\label{second}
\end{equation}

\noindent
where we have defined the auxiliary variables:
\begin{eqnarray}
U_{T} & = & \sum_{ m \alpha \beta n}
\left[\hat G_{jm}^{R}
\hat T_{m \alpha}
(\hat g_{\alpha \beta}^{A}-\hat g_{\alpha \beta}^{R})
\hat T_{\beta n}
\hat G_{ni}^{A}-
\right. \nonumber \\
&& \left. \hat G_{im}^{R} \hat T_{m \alpha}
(\hat g_{\alpha \beta}^{A}-\hat g_{\alpha \beta}^{R})\hat T_{\beta n}
\hat G_{nj}^{A} \right],
\end{eqnarray}
\noindent and
\begin{eqnarray}
U_{S} & = & \sum_{\alpha m} \left[\hat G_{j\alpha}^{R} \hat T_{\alpha m}
(\hat g_{mi}^{A}-\hat g_{mi}^{R})+
 (\hat g_{jm}^{A}-\hat g_{jm}^{R}) \hat T_{m \alpha}
\hat G_{\alpha i}^{A}-
\right. \nonumber \\
& & \left. \hat G_{i\alpha}^{R} \hat T_{\alpha m}
(\hat g_{mj}^{A}-\hat g_{mj}^{R})-
 (\hat g_{im}^{A}-\hat g_{im}^{R}) \hat T_{m \alpha}
\hat G_{\alpha j}^{A} \right]
\end{eqnarray}

The term $(\hat g_{ji}^{+-} - \hat g_{ij}^{+-})$ in (\ref{second}) gives a zero
contribution
to the current because it corresponds
to the current between sites $i$ and $j$ inside the metal in the absence of
tip-sample coupling,
which obviously must be zero.

Let us work now with the term associated to the tip, $U_{T}$.
This term can be simplified by using the well known
relation between advanced and retarded Green's functions:
$\hat G_{nm}^{A}=(\hat G_{mn}^{R})^{\dagger}$.
The real matrix $\hat T_{m \alpha}$ is just equal to
$\hat T_{\alpha m}^{\dagger}$. Hence:

\begin{eqnarray}
\hat G_{im}^{R} \hat T_{m \alpha}
(\hat g_{\alpha \beta}^{A}-\hat g_{\alpha \beta}^{R})\hat T_{\beta n}
\hat G_{nj}^{A} & = & \nonumber \\
 -\left[\hat G_{jn}^{R} \hat T_{n \beta} (\hat g_{\beta \alpha}^{A}-\hat
g_{\beta \alpha}^{R})
\hat T_{\alpha m} \hat G_{mi}^{A}\right]^{\dagger} & &
\end{eqnarray}

So $U_{T}$ can be expressed as the real part of a matrix:

\begin{equation}
U_{T}=2 \Re \sum_{m \alpha \beta n}
\hat G_{jm}^{R}
\hat T_{m \alpha}
(\hat g_{\alpha \beta}^{A}-
\hat g_{\alpha \beta}^{R})\hat T_{\beta n} \hat G_{ni}^{A}
\label{ut}
\end{equation}

The retarded and advanced Green's
functions for the interacting system can
further
be obtained from a
Dyson-like equation that uses the Green's functions of the
uncoupled parts of the system $\hat g^{R}$/$\hat g^{A}$ and the
coupling term $\hat \Sigma^{R}$/ $\hat \Sigma^{A}$
(equation \ref{eqgdyson}):

\begin{equation}
\hat G^{R,A} = \hat g^{R,A} + \hat g^{R,A} \hat \Sigma^{R,A} \hat G^{R,A}.
\end{equation}

If the values of the coupling
matrix $\hat T_{\alpha m}$ are much smaller than hopping
terms inside
the metal (as it is the case in tunneling conditions) we can work
in the lowest order perturbation theory and approximate
$\hat G^{R}_{jm}$ and $\hat G^{A}_{ni}$
simply by $\hat g^{R}_{jm}$ and $\hat g^{A}_{ni}$, respectively.
Moreover,
we can further simplify the expression for $U_{T}$ by relating
$( \hat g_{\alpha \beta}^{A} - \hat g_{\alpha \beta}^{R} )$ with
the density of states matrix at the tip $\hat \rho_{\alpha \beta}$
by the equation
$\hat g_{\alpha \beta}^{A} - \hat g_{\alpha \beta}^{R} = 2 \pi i \hat
\rho_{\alpha \beta}$:

\begin{equation}
U_{T}=4 \pi  \Im \sum_{m \alpha \beta n} \left[\hat g_{jm}^{R}
\hat T_{m \alpha} \hat \rho_{\alpha \beta} \hat T_{\beta n}
\hat g_{ni}^{A}\right]
\end{equation}

The term associated with the sample, $U_S$,
can also be written as the real part
of a matrix using similar arguments leading to (\ref{ut}):

\begin{eqnarray}
U_{S} & = & 2 \Re \sum_{m \alpha} \left[\hat G_{j \alpha}^{R}
\hat T_{\alpha m}
\hat g_{1m}^{A}-\hat g_{jm}^{R} \hat T_{m \alpha} \hat G_{\alpha i}^{A}
\right. \nonumber \\ & & + \left.
\hat g_{jm}^{A} \hat T_{m \alpha} \hat G_{\alpha i}^{A}-\hat G_{j \alpha}^{R}
\hat T_{\alpha m} \hat
g_{mi}^{R}\right]
\end{eqnarray}

Using the Dyson-like equation for the Green's functions of
the interacting system and again working in the
lowest order of perturbation theory:

\begin{equation}
\hat G^{R}_{j \alpha}= \sum_{\beta n} \hat g^{R}_{jn} \hat T_{n \beta} \hat
g^{R}_{\beta \alpha}
\end{equation}

\begin{equation}
\hat G^{A}_{\alpha i}= \sum_{\beta n} \hat g^{A}_{\alpha \beta}
\hat T_{\beta n}
\hat g^{A}_{ni}
\end{equation}

The term associated with the sample $U_S$ can then be expressed as:

\begin{eqnarray}
U_{S} \! &=& 2 \Re \!\!\! \sum_{m \alpha \beta n} \!\! \left[
\hat g_{jm}^{A}
\hat T_{m \alpha}
\hat g^{A}_{\alpha \beta}
\hat T_{\beta n} \hat g_{ni}^{A}-\hat g_{jm}^{R}
\hat T_{m \alpha}
\hat g^{R}_{\alpha \beta}
\hat T_{\beta n} \hat g_{ni}^{R}\right]\!- \nonumber \\
&& 4 \pi \Im \sum_{m \alpha \beta n} 
(\hat g_{jm}^{R}
\hat T_{m \alpha}
\hat \rho_{\alpha \beta}
 \hat T_{\beta n} \hat g_{ni}^{A} 
\end{eqnarray}

It is easy to demonstrate that the first term of this equation is
zero because it is the real part of the difference of two magnitudes,
one being the complex conjugate of the other.
Hence $U_S$ can be shortened to:

\begin{equation}
U_{S}= - 4 \pi  \Im \sum_{m \alpha \beta n}\hat g_{jm}^{R}
\hat T_{m \alpha} \hat \rho_{\alpha \beta} \hat T_{\beta n}
\hat g_{ni}^{A}
\end{equation}

So the rest between
$\hat G_{ji}^{+-}$ and  $\hat G_{ij}^{+-}$ is finally
written:

\begin{equation}
\hat G_{ji}^{+-} -
\hat G_{ij}^{+-} = 4 \pi (f_{T} - f_{S})  \Im \sum_{m \alpha \beta n}
\hat g_{jm}^{R} \hat T_{m \alpha} \hat
\rho_{\alpha \beta} \hat T_{\beta n} \hat g_{ni}^{A}
\end{equation}

Assuming
zero temperature, we obtain
the current between two sample sites $i$ and $j$
in the sample as an integral over a window of energies
ranging from the Schottky Barrier Height ($eV_o$) up to
the applied voltage of the imaginary part of the trace of
a product of matrices:

\begin{equation}
J_{ij}(V) = \frac{4e}{\hbar} \Im \int_{eV_o}^{eV} \! \! \!
Tr \; \sum_{m \alpha \beta n} \left[ \hat T_{ij}
\hat g_{jm}^{R}
\hat T_{m \alpha}
\hat
\rho_{\alpha \beta}
\hat T_{\beta n} \hat g_{ni}^{A} \right] dE
\label{apfinal}
\end{equation}

This is equation (\ref{jreal}), that can be considered as
our starting point.

\end{document}